\documentclass[prb,twocolumn,showpacs,superscriptaddress]{revtex4}
\usepackage[dvips]{graphicx}
\usepackage{dcolumn}
\usepackage{bm}
\usepackage{amssymb}
\usepackage{textcomp}

\DeclareGraphicsExtensions{eps}

\begin{document}

\preprint{APS/123-QED}
\title{Different effects of Ni and Co substitution on the transport properties of BaFe$_2$As$_2$ }

\author{A. Olariu}
\email{areta.olariu@cea.fr}
\author{F. Rullier-Albenque}
\author{D. Colson}
\author{A. Forget}

\affiliation{Service de Physique de l'Etat Condens\'e, Orme des Merisiers, IRAMIS, CEA-Saclay (CNRS URA 2464), 91191 Gif sur Yvette Cedex, France}

\date{\today}

\begin{abstract}
We report resistivity and Hall effect results on Ba(Fe$_{1-x}$Ni$_x$)$_2$As$_2$ and compare them with those in Ba(Fe$_{1-x}$Co$_x$)$_2$As$_2$. The Hall constant $R_H$ is negative for all $x$ values from 0.01 to 0.14, which indicates that electron carriers dominate the transport in both the magnetic and the paramagnetic regimes. We analyze the data in the framework of a two-band model. Without any assumption on the number of carriers, we show that the electron resistivity can be estimated with good accuracy in the low temperature paramagnetic range. Although the phase diagrams of the two families are very similar with respect to the extra electrons added in the system, we find that the transport properties differ in several aspects. First, we evidence that the contribution of holes to the transport is more important for Ni doping than for Co doping. Second, Ni behaves as a stronger scatterer for the electrons, as the increase of residual electron resistivity $\Delta \rho_e/x$ is about four times larger for Ni than for Co in the most doped samples.
\end{abstract}

\maketitle

\section{Introduction}

The newly discovered iron pnictides~\cite{Kamihara08} are attracting great interest, stimulated by their high critical temperatures, and the resemblance with the well-studied cuprates. Regardless of similar phase diagrams, important differences exist between the two families. Iron pnictides are metallic with a multiband electronic structure, while cuprates are strongly correlated Mott insulators with a single band behavior. The coexistence of magnetism and superconductivity observed at nanoscale level for many Fe-based systems marks another important difference.

Among all the iron pnictides, the Ba-based 122 family is so far the most studied, because superconductivity can be induced by
doping with electrons~\cite{Sefat08,Chu09}, holes~\cite{Rotter08}, or by chemical~\cite{Ren09-PRL,Jiang09} or mechanical pressure~\cite{Alireza09}. The shapes of the phase diagrams are similar and believed to be induced by the interplay between magnetic and electronic properties. The electronic band structure was calculated using the local density approximation (LDA) and shown to be semimetal-like, dominated by five Fe $d$ states at the Fermi energy. The Fermi surface consists of small hole and electron pockets, containing few carriers~\cite{Singh09}. The parent compound shows no superconductivity and presents an antiferromagnetic long range order at low temperature~\cite{Huang08}.

The effect of electron and hole doping on the electronic structure is often considered in a rigid-band approximation in which the topology of the Fermi surfaces does not evolve with doping, while the Fermi level is shifted. Thus the size of the pockets changes continuously with doping \cite{Sefat08}. This image was confirmed by angle-resolved photoemission spectroscopy (ARPES) experiments for electron doping on Ba(Fe$_{1-x}$Co$_x$)$_2$As$_2$~\cite{Sekiba09,Brouet09} and for hole doping on Ba$_{1-x}$K$_x$Fe$_2$As$_2$~\cite{Liu08}. However, recent density functional theory (DFT) calculations suggest that Co and Ni do not carrier-dope, but instead are isovalent of Fe and behave as random scatterers~\cite{Wadati10}. The effect of substitution would rather be to wash out certain parts of the Fermi surface, and to destabilize the magnetism in favour of superconductivity.

Transport is a well suited technique to study the electronic structure and its modification with doping. Although a large number of studies have been devoted to the 122 family, many issues still need to be clarified. In the parent compound BaFe$_2$As$_2$, for which the number of holes and electrons are equal, the Hall constant is negative at all temperatures, which shows that electrons dominate the transport. Holes appear to be highly scattered, and therefore not directly visible in the transport properties. The same behavior was observed in Ba(Fe$_{1-x}$Co$_x$)$_2$As$_2$~\cite{Rullier-Albenque09,Katayama09,Fang09} and Sr(Fe$_{1-x}$Ni$_x$)$_2$As$_2$~\cite{Butch10}. A positive Hall constant is measured only in hole doped systems Ba$_{1-x}$K$_x$Fe$_2$As$_2$~\cite{Luo09} and Ba(Fe$_{1-x}$Cr$_x$)$_2$As$_2$~\cite{Sefat09-Cr}, as well as in Ba(Fe$_{1-x}$Ru$_x$)$_2$As$_2$, for which Ru is isovalent of Fe~\cite{Rullier-Albenque10}.

Further insight in the transport properties is obtained by doping with Ni at the Fe site in the Ba family. The phase diagram obtained from resistivity measurements is similar to the other 122 families~\cite{Li09,Canfield09,Ni10}. It was shown that good matching is obtained with the phase diagram of Ba(Fe$_{1-x}$Co$_x$)$_2$As$_2$ if the $x$-axis is rescaled by a factor of 2~\cite{Canfield09}. This is in line with the rigid-band image of the doped electronic structure, for which Ni provides two extra-electrons to the electronic bands, while Co gives only one.

In this paper we present a systematic study of the resistivity and the Hall effect in Ba(Fe$_{1-x}$Ni$_x$)$_2$As$_2$ with $x$ ranging from 0.01 to 0.14 and we compare our results with those obtained previously in Ba(Fe$_{1-x}$Co$_x$)$_2$As$_2$~\cite{Rullier-Albenque09}. The quantitative analysis of the data is performed in a two-band model. We show that it is possible to get relevant information on the respective resistivities of the electrons and holes \textit{without making any assumption on the effective number of carriers}. For these two families, we find that the electron band dominates the transport properties in the whole doping range and at all temperatures. In particular, in the paramagnetic state, the hole resistivity is estimated to be at least 2-5 times larger than that of electrons for Ni doping, and even more for Co doping. We also find that Ni induces more scattering than Co in the electron band. For the most overdoped sample, the electron mobility is reduced by a factor of 2 when Ni, rather than Co, is substituted to Fe.

\section{Synthesis and sample preparation} Single crystals of Ba(Fe$_{1-x}$Ni$_x$)$_2$As$_2$ with $x$ ranging from 0.01 to 0.14 were grown using the self-flux method. High purity Ba pieces, FeAs and NiAs powders, mixed together in the ratio 1: 4-$x$: $x$, were put in alumina crucibles and sealed in evacuated quartz tubes. The mixture was held at 1180$^\circ$C for 4 h. It was then cooled down to 1000$^\circ$C at 6$^\circ$C/h, then down to room temperature at 100$^\circ$C/h. Single crystals were extracted mechanically from the flux. Chemical analysis was performed on several crystals of each batch with Camebax SX50 electron microprobe in several spots of the surfaces. This technique gives a precision of typically 0.4\% on the Ni content.

Resistivity and Hall effect measurements were subsequently performed on the same crystals. Single crystals were cleaved to thickness smaller than 30~$\mu$m in order to ensure a good homogeneity. Samples are squarelike with sides ranging from 0.3 to 0.7~mm. Contacts were made using silver epoxy. For each Ni content, at least two different samples were analyzed. Resistivity was measured using the van der Pauw technique~\cite{vanderPauw58}. We found very good reproducibility of the absolute value of the resistivity and of the Hall number, with $\sim$10\% variation from one sample to another for the same Ni content.

\section{Resistivity measurements}

\begin{figure*}
\includegraphics[width=\linewidth]{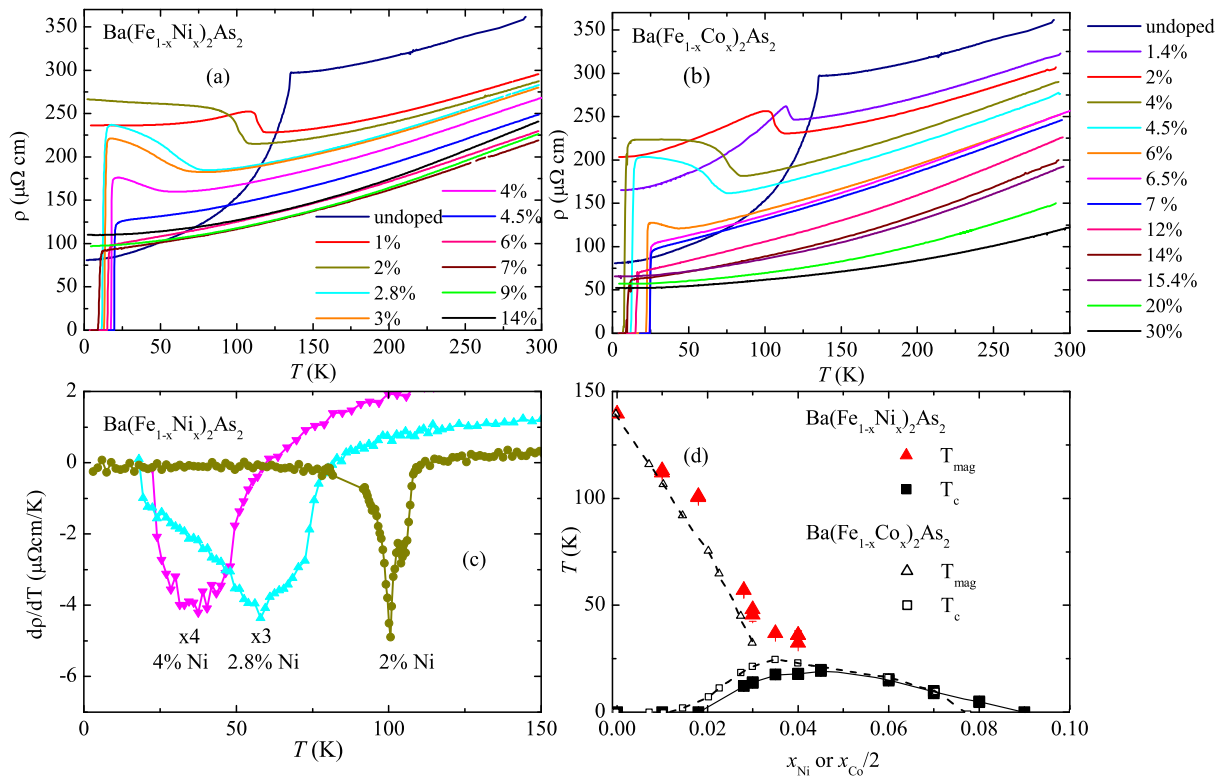}
\caption{(Color online) Temperature dependence of the resistivity for a) Ba(Fe$_{1-x}$Ni$_x$)$_2$As$_2$ and b) Ba(Fe$_{1-x}$Co$_x$)$_2$As$_2$. Same colours were used as much as possible for respective $\rho(T)$ curves corresponding to equivalent electron doping, that is concentrations of Ni and Co of $2x$ and $x$ respectively. c) Derivative of the resistivity as a function of the temperature for underdoped Ni-samples. d) Phase diagram indicating $T_{\rm{mag}}$ and $T_c$ for the two families. In order to allow comparison between the two families, Co content was divided by 2. Lines are guides for the eye.}
\label{Res}
\end{figure*}

The temperature dependence of the resistivity is shown in Fig.~\ref{Res}a) for the Ba(Fe$_{1-x}$Ni$_x$)$_2$As$_2$ family and in Fig.~\ref{Res}b) for Ba(Fe$_{1-x}$Co$_x$)$_2$As$_2$. Superconductivity occurs for Ni content in the range 0.02 to 0.09. The resistivity drops to zero at $T_c$ on less than a 0.5~K range, which shows that samples are very homogeneous.  The optimum value of $T_c$ is 19.5$\pm0.5$~K, obtained for $x=0.045$.
For $x\leq 0.04$, samples present a strong increase of the resistivity as $T$ decreases in the intermediate temperature range around 100~K . Similar behavior was observed for other 122 families, in particular for Ba(Fe$_{1-x}$Co$_x$)$_2$As$_2$ for $x\leq0.06$~\cite{Rullier-Albenque09}. This signals the occurrence of the structural and magnetic phase transitions. As shown by x-rays and neutron diffraction measurements~\cite{Pratt09}, a good determination of the respective transition temperatures can be done by considering the temperature derivative of the resistivity $d\rho/dT$ versus $T$ (see Fig.~\ref{Res} c) in which the minimum of $d\rho/dT$ is associated to the magnetic transition $T_{\rm{mag}}$ while the broad feature at slightly higher temperature is due to the structural transition.

The phase diagram ($T_c$ and $T_{\rm{mag}}$) obtained from the resistivity data is shown in Fig.~\ref{Res} d) and is compared to that of Ba(Fe$_{1-x}$Co$_x$)$_2$As$_2$ indicated as empty symbols and dotted lines. In this plot, the Co concentration was divided by 2 to take into account that only one electron is added to the electronic bands in this case. Our results are in very good agreement with those reported in Ref~\cite{Canfield09,Ni10}. There is a slight mismatch between the shapes of the of the Ni and Co phase diagrams. In particular, the magneto-structural transition persists to a slightly larger doping for Ni than for Co substitution. This might explain why the maximum $T_{c}$ is also smaller in Ni doped compounds, since one observes that the two superconducting domes merge on a single curve once the magnetic state is totally suppressed in Ni doped compounds. Let us stress here that the scaling between these two phase diagrams as a function of the extra electrons added to the band structure is a strong indication that the main effect of Co and Ni is to electron dope the system. Moreover, we see hereafter that Ni is a stronger scatterer than Co. These two observations are in contradiction with the suggestion done in Ref\cite{Wadati10} which attributes the main effect of Co or Ni substitution to a scattering effect.

As seen in Figs.~\ref{Res} a) and b), tiny substitution of either Ni or Co has a strong effect on the resistivity in the paramagnetic regime. A drastic drop of the absolute value of $\rho$ occurs for the smallest concentrations considered in these studies. The resistivity then decreases progressively with $x$, and the absolute resistivity values remain very similar for the Ni- and Co-doped samples corresponding to the same effective doping in the underdoped regime. However, a different behavior is observed at larger doping. For $x_{\rm{Ni}}\geq 0.07$, the absolute value of the resistivity \emph{increases} as more nickel is added to the structure. On the contrary, $\rho$ continues to gradually decrease with $x$ for Co doping. Consequently the resistivity at $x_{\rm{Co}}=0.3$ is about twice smaller than for equivalent Ni doping. In conclusion, good matching of the resistivity data is obtained at low doping, but discrepancies appear at larger doping.

\begin{figure}
\includegraphics[width=\linewidth]{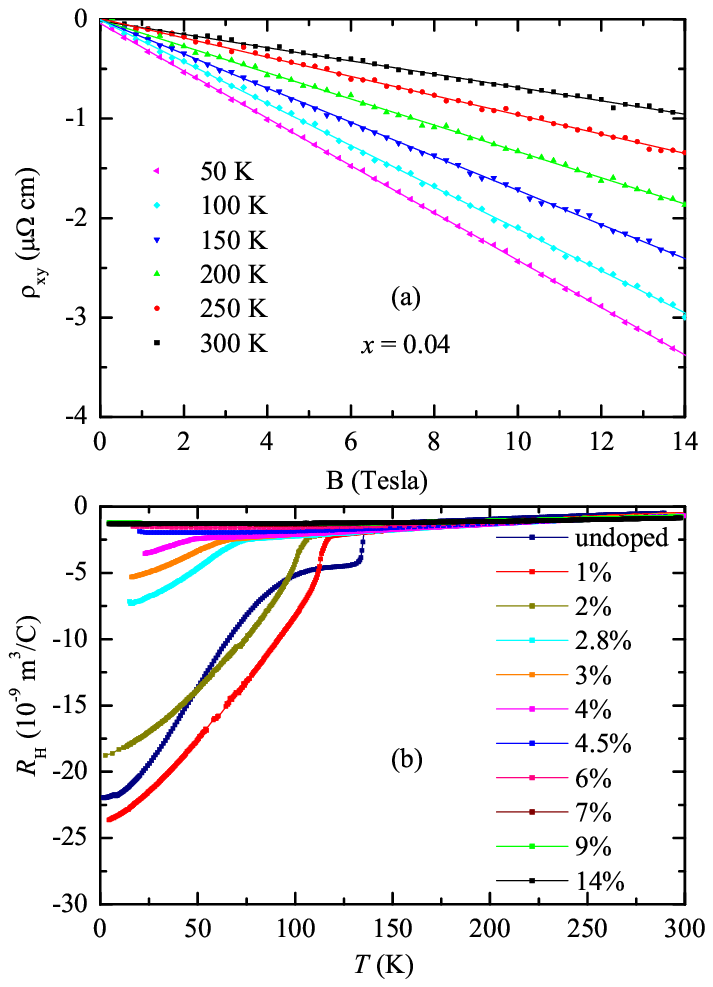}
\caption{(Color online) a) The Hall resistivity at different temperatures for $x$=0.04. Continuous lines are linear fits. b) Temperature dependence of the Hall coefficient $R_H$ obtained for the Ba(Fe$_{1-x}$Ni$_x$)$_2$As$_2$ family.}
\label{RH}
\end{figure}

\section{Hall effect} In order to obtain more insight into the transport properties of the system, we performed Hall effect measurements in the ($ab$) plane. Single crystals were placed in either an 8~T or a 14~T magnetic field perpendicular to the sample surface. In the paramagnetic state, the Hall resistivity $\rho_{xy}$ is linear in magnetic field and has a negative slope at all temperatures, as shown in Fig.~\ref{RH} a) for $x=0.04$. This allows us to determine the Hall constant $R_H$ versus temperature and doping. Some departures from linearity are observed in the magnetic phase and $R_{H}$ is then estimated from the values obtained at the maximal field. The evolution of $R_{H}$ is shown in Fig.~\ref{RH} b). $R_H$ is negative at all temperatures and for all doping values, which indicates that the electronic contribution is dominant. A very pronounced decrease of the Hall constant is observed below the magnetic transition at low doping, similarly to Ba(Fe$_{1-x}$Co$_x$)$_2$As$_2$. This is certainly induced by the the strong reduction of the number of carriers due to the gapping of the Fermi surfaces~\cite{Hu08}.

\begin{figure*}
\includegraphics[width=\linewidth]{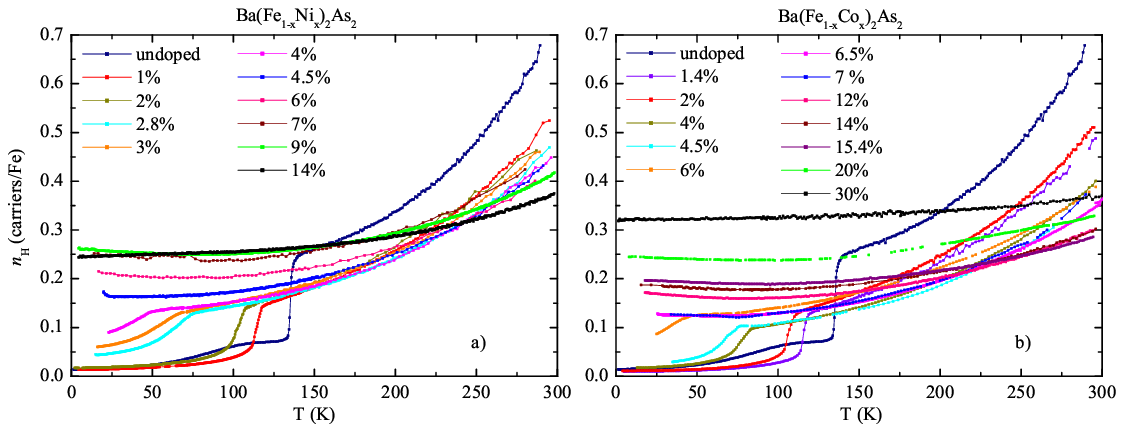}
\caption{(Color online) Temperature dependence of the Hall number $n_H$ of a) the Ba(Fe$_{1-x}$Ni$_x$)$_2$As$_2$ family and b) the Ba(Fe$_{1-x}$Co$_x$)$_2$As$_2$ family.}
\label{nH}
\end{figure*}

We now concentrate on the paramagnetic state.
In order to better visualize the temperature variation of the Hall constant of the different samples,  it is more relevant to consider here the Hall number $n_H$=-1/$eR_H$. The temperature dependence of $n_H$ is presented in Fig.~\ref{nH} a), together with the curves obtained for the Co-family, shown in Fig.~\ref{nH} b). Similarly to the resistivity, a drastic drop of the Hall number occurs between the pure and the lightly doped $x_{\rm{Ni}}=0.01$ system. As indicated in the Co case~\cite{Rullier-Albenque09}, this behavior might signal a profound modification of the Fermi surface upon doping and could be related to the Lifshitz transition evidenced by ARPES in Ba(Fe$_{1-x}$Co$_x$)$_2$As$_2$~\cite{Liu10}.

For $T \gtrsim 150$K, the Hall number displays a  strong increase with temperature for underdoped samples with a factor of $\sim3$ between 150 and 300~K. For overdoped samples, a very broad minimum occurs below $\sim100$~K, then a moderate increase at high $T$. Similar temperature dependences are observed for  Ba(Fe$_{1-x}$Co$_x$)$_2$As$_2$~\cite{Rullier-Albenque09,Katayama09,Fang09} as reported in Fig.~\ref{nH} b).

\begin{figure}
\includegraphics[width=\linewidth]{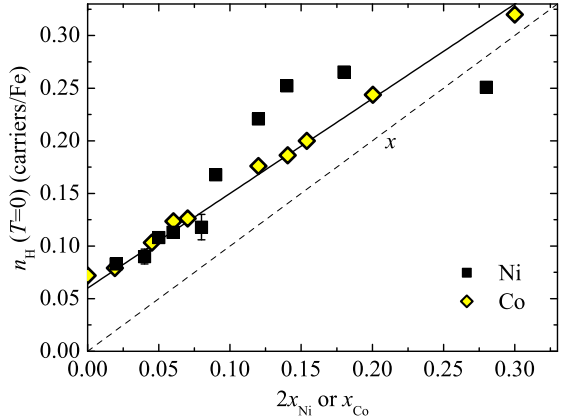}
\caption{(Color online) Hall number extrapolated to $T=0$ for Ni and Co-doped samples. Ni concentration was multiplied by a factor of two. The continuous line is a linear fit to the Co-data. }
\label{nH0}
\end{figure}

Let us focus first on the zero temperature variation of $n_{H}$ with doping. For underdoped samples, $n_{H}$ is estimated by extrapolation at $T=0$ from a parabolic fit of the data in the paramagnetic regime. This determination results in some uncertainty in $n_{H}(T=0)$, especially at very low doping content. Results are shown in Fig.~\ref{nH0}, together with similar results for Ba(Fe$_{1-x}$Co$_x$)$_2$As$_2$. In order to compare the two families, the Ni concentration was multiplied by a factor of 2. In the case of Co, $n_H(T=0)$ varies linearly with doping in the range $0.02\leq x \leq 0.20$, with a slope of slightly less than 1~carrier/Fe~\cite{Rullier-Albenque09}. In order to account for the transport properties, it was suggested that only electrons contribute, so that Hall number equals the number $n_e$ of electron carriers. This was confirmed by ARPES measurements performed on the same samples in which $n_e$ is estimated from the size of the electron pockets~\cite{Brouet09}. One can note that the number of electrons and holes deduced from ARPES measurements are considerably smaller than those given by LDA calculations. This discrepancy between experiment and calculations has been already pointed out in different reports~\cite{Ortenzi09,Benfatto10-condmat}. In BaFe$_2$As$_2$ for instance, $n_e=n_h=0.06\pm0.02$ from ARPES, and 0.15 from LDA~\cite{Fang09}.

In the case of Ni, the variation of $n_H$ with $x$ is non-linear. In the underdoped regime, rather good matching is obtained for the two families. At large doping, for $x_{\rm{Ni}}$=0.14, $n_H=0.25\approx 2x_{\rm{Ni}}$; that is, no holes are left in the system. At intermediate doping level $0.08 \leq x_{\rm{Ni}} < 0.14$, very different values are found with respect to Co. $n_H$ is much larger than expected if only electrons are taken into account. Therefore, unlike for Ba(Fe$_{1-x}$Co$_x$)$_2$As$_2$, transport is not totally dominated by electrons. As we see below, this effect could be accounted for in a two-band model by considering the contribution of holes to the transport.

\section{Analysis of the transport results}

\subsection{The two-band model}

In order to interpret the resistivity and Hall effect data in the paramagnetic state, we consider as a starting point the two-band model, for which the Hall constant is given by the formula:

\begin{equation}
eR_H=-\frac{1}{n_H}=\frac{1}{\sigma^2}\Big(\frac{\sigma_h^2}{n_h}-\frac{\sigma_e^2}{n_e}\Big).
\end{equation}

Here $n_e$ and $n_h$ are the concentrations of electrons and holes in the system, while $\sigma$ is the total conductivity, given by the contribution of both electrons and holes: $\sigma=1/\rho=\sigma_e+\sigma_h$. Since the first term of Eq.~(1) is positive, and $\sigma_e/\sigma \leq 1$, it results that $n_H$ is an upper limit for $n_e$; that is, $n_e \leq n_H$ for all doping values. The charge conservation writes for Ni-doping:

\begin{equation}
n_e=n_h+2x,
\end{equation}

assuming that each Ni atom gives two electrons to the bands. Note that in the case of Co-doping, the charge conservation implies $n_e=n_h+x$.

We first analyze results obtained at large doping, for $x_{\rm{Ni}}=0.14$. In this case $n_H \approx 2x_{\rm{Ni}}$ for $T<180$~K, and only electrons contribute to the transport, so that $n_e \approx n_H$ and $\rho_e \approx \rho$. The resistivity is well fitted by a parabolic law: $\rho\approx\rho_e = \rho_e(T=0)+B T^2$, which is the signature of a Fermi-liquid behavior. The residual resistivity, $\rho_e(T=0)\sim 110\  \mu \Omega$cm is about twice larger than for equivalent Co-doping, indicating that Ni-induced electron scattering is much stronger than for Co-doping.

For $x_{\rm{Ni}}<0.14$, $2x_{\rm{Ni}} < n_H$ for all $T$ in the paramagnetic regime. In a two-band model, Eq. (1) tells us that the contribution of holes cannot be neglected. So four unknown variables: $n_e$, $n_h$, $\sigma_e$ and $\sigma_h$ appear in the three relationships given above. In order to disentangle the electron and hole contributions, the number of carriers at each doping value is required. No such data exist so far for Ba(Fe$_{1-x}$Ni$_x$)$_2$As$_2$. In the following, we show how we can get rather good estimates for $\rho_e$ in a two band model without any assumptions on the number of carriers. This allows us to compare the results obtained with Ni and Co dopings by the same analysis.

\subsection{An estimate of the electron resistivity}

\begin{figure}
\includegraphics[width=\linewidth]{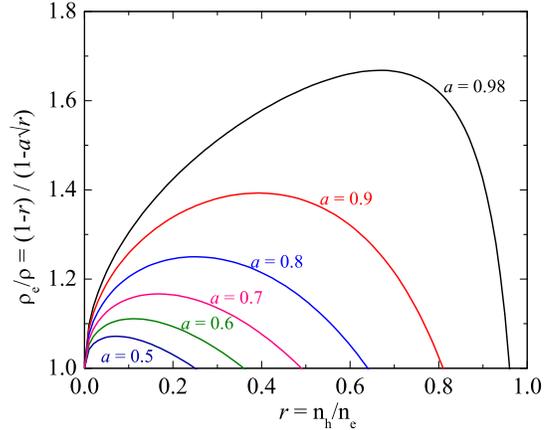}
\caption{(Color online) Ratio $\rho_e/\rho$ for different values of $a=\surd(1-2x/n_H)$. For each sample, $a$ is estimated on the basis of Hall effect measurements. }
\label{f(r)}
\end{figure}

We will show now that it is possible to extract interesting information from the resistivity and Hall effect data within the two-band model but without doing any assumption on the doping evolution of the number of carriers. For the electron resistivity, a simple calculation shows that $\rho_e$ can be expressed as a function of the total resistivity, according to the formula:

$$
\rho_e(T)=\frac{1-r}{1-a(T) \surd{r}} \rho(T)
$$

\begin{figure*}
\includegraphics[width=\linewidth]{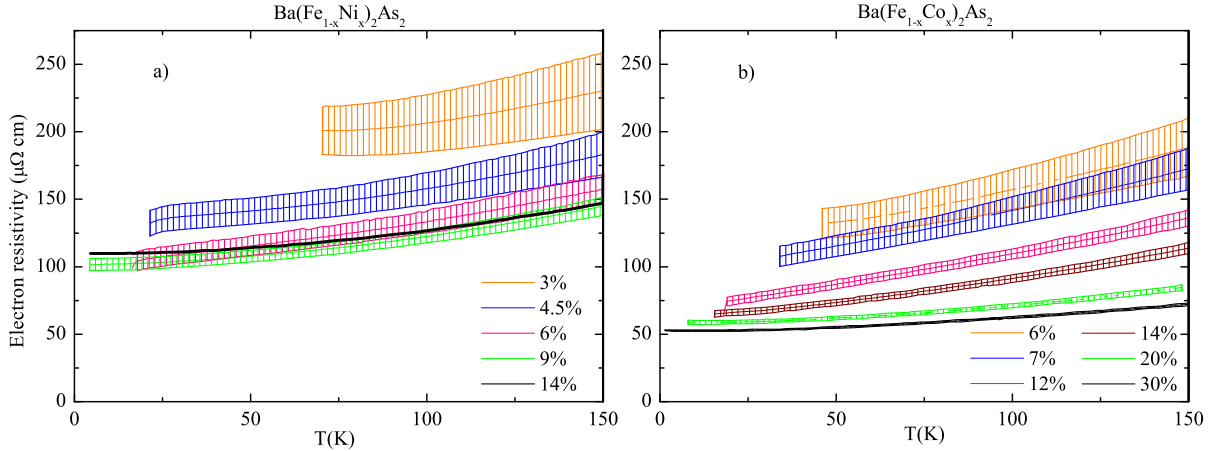}
\caption{(Color online) Temperature dependence of the electron resistivity estimated in a two-band model for a) Ba(Fe$_{1-x}$Ni$_x$)$_2$As$_2$ and b) Ba(Fe$_{1-x}$Co$_x$)$_2$As$_2$. The error bars are given by the lower and upper limit of $\rho_e$ in a two-band model, with no assumption on the number of carriers.}
\label{ResE(T)}
\end{figure*}

\begin{figure}
\includegraphics[width=\linewidth]{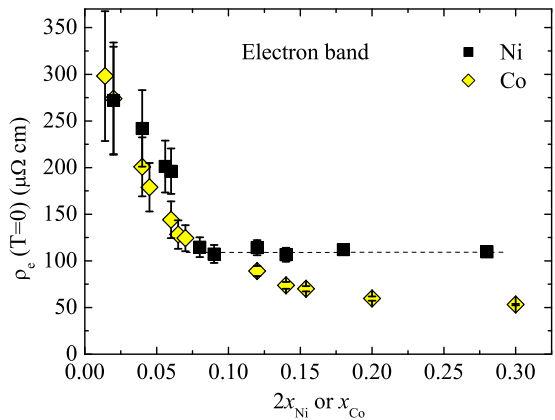}
\caption{(Color online) Electron resistivity at $T=0$ in a two-band model, for Ni and Co-doped samples. No assumption is made regarding the number of carriers and its variation with doping. Error bars are given by the extreme values estimated within this analysis. The fact that $\rho_{e}(T=0)$ saturates at large dopings for Ni doping indicates that the scattering rate of electrons increases in this case whereas it is found almost independent of Co doping (see text).}
\label{Modele2bandesRho0}
\end{figure}

\begin{figure}
\includegraphics[width=\linewidth]{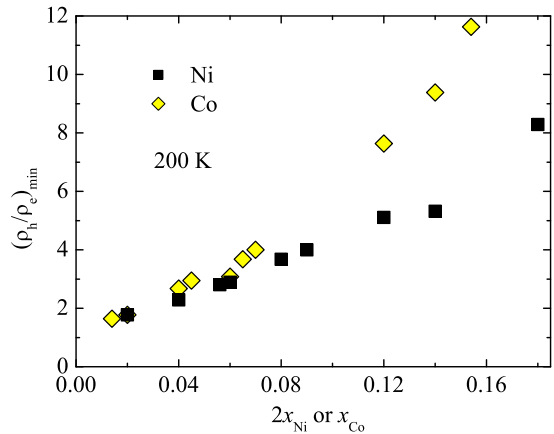}
\caption{(Color online) Minimum value of $\rho_h/\rho_e$ estimated at 200~K for Ni and Co-doped systems. Similar curves are obtained at all temperatures in the paramagnetic range. }
\label{ResHResE}
\end{figure}

Here $r=n_h/n_e \leq 1$ is unknown, while $a(T)=\surd (1-2x/n_H(T))$ ($a<1$) can be calculated for every sample with $x_{\rm{Ni}}<0.14$ from Hall effect measurements. For each value of $a$, the function $f(r)=\rho_e/\rho=(1-r)/(1-a\surd{r})$ is a dome-shaped curve, as shown in Fig.~\ref{f(r)}. It varies from 1 to a maximum value that decreases with $a$. For every sample, the range of variation of the ratio $\rho_e/\rho$ is obtained in this way at each temperature. We estimate an average value and error bars given by the lower and upper limits of $f(r)$. For a given concentration, the best estimate of $\rho_e/\rho$  is therefore obtained at low $T$, where $a(T)$ is minimum. Very good accuracy is obtained in the overdoped regime, where $a$ approaches 0.5 at low temperature, so that $f(r)$ is very close to 1. By similar calculations it is easily shown that $d\rho_e/dT  \geq 0$, hence the resistivity of electrons always increases with $T$, indicating a metallic character for the electron band.

The electron resistivity estimated in this way is shown versus temperature up to 150~K in Fig.~\ref{ResE(T)} a) for Ba(Fe$_{1-x}$Ni$_x$)$_2$As$_2$ and Fig.~\ref{ResE(T)} b) for Ba(Fe$_{1-x}$Co$_x$)$_2$As$_2$. Only a few concentrations are shown, for sake of clarity. The differences between the two families in the overdoped regime are clearly visible: $\rho_e$ saturates for Ni-doping, while it progressively decreases for Co-doping. Considerably larger error bars are obtained in the case of Ni-doping with respect to Co-doping. This reflects the fact that it is possible to neglect the hole contribution in the latter case, while this has to be taken into account for Ni doping.

The residual resistivity of electrons for $T \rightarrow 0$ is shown in Fig.~\ref{Modele2bandesRho0} for both Ni and Co-doped samples. The extrapolation to $T=0$ is done by a parabolic fit of $\rho$ in the low-$T$ paramagnetic range. While $\rho_{e}(T=0)$ evolves similarly for Co and Ni doping in the underdoped regime, a pronounced difference appears at larger doping. In the case of Co-doping, $\rho_e(T=0)$ decreases progressively with $x_{\rm{Co}}$. Instead of this, a saturation of $\rho_e(T=0)$ is clearly observed for Ni doping once $x \geq 0.04$. Assuming $n_e=n_H(T=0)$ for Co doping indicates that the residual scattering rate of electrons is rather independent of Co-content, as found previously in Ref~\cite{Rullier-Albenque09}. On the contrary it strongly increases with Ni substitution. In particular, for the largest doping studied here, the increase of scattering rate is two times larger for Ni than for Co, that is a factor of 4 with respect to the impurity content.

While $\rho_e$ can be estimated with rather good accuracy using this approach, hole resistivity is very sensitive to the number of carriers in the system. The value of $\rho_h$ can be estimated directly from:

$$
\frac{1}{\rho_h}= \frac{1}{\rho} -\frac{1}{\rho_e}
$$

The ratio $\rho_h/\rho$ presents very large variation with $r$. Thus an accurate determination of $\rho_h$ and of $d\rho_h/dT$ cannot be performed  without the precise knowledge of the number of carriers. In particular, the derivative $d\rho_h/dT$ is found to be either positive or negative, depending on $r$. It is therefore not possible to discern between a semiconducting or a metallic character for the hole band. However, it is possible to make an estimate of the minimum values of the ratios $\rho_h/\rho$ and $\rho_h/\rho_e$. It is easily shown that the first ratio is larger than 2, so that $\rho_h \geq 2\rho$ for any temperature and doping. The minimal values of $\rho_h/\rho_e$ are shown at $T=200$~K in Fig.~\ref{ResHResE} for Ni and Co-doped samples. Similar curves are obtained at all temperatures, though the values slightly decrease with $T$. For $x_{\rm{Ni}} \geq 0.04$, the hole resistivity is at least 3.5 larger than the electron resistivity at 200~K. Even larger values are obtained for Co-doping. This is consistent with the observation that only electrons contribute to the transport from the previous study on
Ba(Fe$_{1-x}$Co$_x$)$_2$As$_2$~\cite{Rullier-Albenque09}.

\section{Discussion and Conclusion}
Our resistivity measurements on Ba(Fe$_{1-x}$Ni$_x$)$_2$As$_2$ confirm the phase diagram previously established~\cite{Ni10}.
Negative Hall constant is found in the whole phase diagram, in both the magnetic and the paramagnetic regimes. This shows that the electron bands dominate the transport properties, similar to what is observed in Ba(Fe$_{1-x}$Co$_x$)$_2$As$_2$ and Sr(Fe$_{1-x}$Ni$_x$)$_2$As$_2$. At large doping, where no holes are left in the system, the Hall number is close to $2x_{\rm{Ni}}$. This confirms the assumption that each Ni atom adds two electrons to the bands and that its main effect is to electron dope the system, in contradiction to what is claimed in Ref\cite{Wadati10}.

In the intermediate doping range we show that both electron and hole contributions must be considered to account for the Hall data. We have presented a general analysis of the transport data within the two-band model which made it possible to have a good estimate of the electron resistivity without any assumption on the number of carriers in the respective bands. In fact we show that $\rho_e$ and $\rho_h$ only depend on the ratio $r=n_h/n_e$. Regardless of the exact value of $r$, the residual value of $\rho_e$ and its temperature dependence can be estimated with rather good accuracy for $x_{\rm{Ni}} \geq 0.04$ and $T \lesssim 200$K. This method allowed us to compare precisely the effect of Co and Ni doping on the transport properties.

Although a very good matching is observed between the phase diagrams of these two systems when $x$ is scaled by a factor of 2, some discrepancies appear in their transport properties. While  it appears that the hole contribution is barely visible in the transport properties of Ba(Fe$_{1-x}$Co$_x$)$_2$As$_2$, we clearly evidence that it must be taken into account in the case of Ba(Fe$_{1-x}$Ni$_x$)$_2$As$_2$. Moreover, a previous analysis led us to suggest that a $T$-dependent variation of the number of electrons should be foreseen to understand the Hall effect in Ba(Fe$_{1-x}$Co$_x$)$_2$As$_2$ \cite{Rullier-Albenque09}. Let us notice that a recent theoretical approach apparently supports the idea that the $T$-dependence of $\rho$ is not solely due to scattering processes but also to a variation of the carrier content~\cite{Benfatto10-condmat}. The approach considered here is thus particularly powerful as it allowed us to extract general information without precluding a possible variation of $n_e$ or $n_h$ with temperature.

It has been recently argued that the linear $T$ dependence of resistivity found near optimal doping for Co doping or isovalent P substitution in place to As might be the signature of spin fluctuations and non Fermi liquid behavior \cite{Doiron09, Kasahara10}. Let us note here that we do also observe a linear temperature dependence of the resistivity for $x_{\rm{Ni}}=0.045$ corresponding to optimal doping. However in this doping range, both electrons and holes contribute to the transport. The analysis performed in section V clearly shows that the electron resistivity is rather quadratic in temperature, indicating that any conclusion from a peculiar $T$ variation of the \textit{total} resistivity must be considered with caution in a multiband system.

An important result of our study is to show that Ni behaves as a much stronger scatterer for electrons than Co. In particular we find that $\Delta\rho_{e}/x$ is about four times larger in the former case at large doping. On the other hand, the $T_{c}$ values appear very similar for $x_{\rm{Ni}} \gtrsim 0.04$ and $x_{\rm{Co}} \gtrsim 0.08$. This suggests that $T_{c}$ is rather insensitive here to the additional scattering induced by Ni substitution. It has been often suggested that the lower $T_{c}$ induced by Co doping in BaFe$_2$As$_2$ compared to K doping might be induced by the presence of dopant atoms into the FeAs layers, which is expected to be detrimental for superconductivity. Recent experiments have shown that the introduction of defects in Ba(Fe$_{1-x}$Co$_x$)$_2$As$_2$, either through the substitution of Fe by Zn~\cite{Li11} or by irradiation~\cite{Nakajima10}, results in strong decrease of $T_{c}$ \cite{}. In this context, it seems to us very intriguing that Ni doping which appears to strongly increase the residual resistivity of electrons has no effect on the superconducting critical temperature.

In conclusion, although a rapid examination of the phase diagrams of Co and Ni doped compounds suggest that these two systems are very similar, our study reveals some differences in their transport properties. This must be related to subtle differences in the evolution of their Fermi surfaces upon these different types of doping. This opens the possibility to study the relationship between the electronic properties in the different bands and superconducting properties, which might help for the understanding of the intra- and interband scattering processes.

\begin{acknowledgments}
Authors wish to thank H. Alloul and P. Bonville for helpful discussions and critical reading of the manuscript and S. Poissonnet (SRMP/CEA) for the
chemical analysis of the samples. This work was supported by the ANR grant "PNICTIDES". A.O. acknowledges financial support from "Triangle de la Physique".
\end{acknowledgments}


\end{document}